# Size Dependent Breakdown of Superconductivity in Ultranarrow Nanowires

Maciek Zgirski, Karri-Pekka Riikonen, Vladimir Touboltsev, and
Konstantin Arutyunov*

*NanoScience Center, Department of Physics, University of Jyväskylä, PB 35,
40014, Jyväskylä, Finland*

**ABSTRACT**

Below a certain temperature $T_c$ (typically cryogenic), some materials lose their electric resistance $R$ entering a superconducting state. Following the general trend toward a large scale integration of a greater number of electronic components, it is desirable to use superconducting elements in order to minimize heat dissipation. It is expected that the basic property of a superconductor, i.e., dissipationless electric current, will be preserved at reduced scales required by modern nanoelectronics. Unfortunately, there are indications that for a certain critical size limit of the order of ~10 nm, below which a "superconducting" nanowire is no longer a superconductor in a sense that it acquires a finite resistance even at temperatures close to absolute zero. In the present paper we report experimental evidence for a superconductivity breakdown in ultranarrow quasi-1D aluminum nanowires.

Zero electric resistance is one of the basic properties of a superconductor. In conventional metallic materials (e.g., tin, lead, indium, aluminum), the dissipationless electric current originates from a particular interaction between electrons and vibrating ion cores forming the crystalline lattice. One might expect that superconductivity should be size independent as long as the so-called electron−phonon interaction is not dramatically modified by the size-related localization such as, e.g., in nanometer clusters comprising finite number of atoms. Of a practical interest are the prospects for utilization of superconducting components in nanoelectronic circuits. A key question here is: can a "superconducting" wire of ~10 nm in diameter (≥ 10 000 atoms within the cross section) sustain a dissipationless electric current? Theoretical predictions and the results of the present experimental study are rather pessimistic: below a certain limit a "superconducting" wire acquires a finite resistance due to quantum fluctuations destroying its superconducting state even at temperatures tending to absolute zero.

Superconductivity is a macroscopically coherent state. In equilibrium the quantum state is described by a single wave function $\Psi = |\Psi| e^{i\varphi}$, where the magnitude $|\Psi|$ can change on scales larger than the superconducting coherence length $\xi$, and the phase $\varphi$ is "locked" all over the whole system, being related to the supercurrent density $j \sim \nabla\varphi$. A superconducting wire can be considered as one-dimensional (1D) if its smallest transverse dimension $\sqrt{\sigma}$ is smaller than the coherence length $\xi$. Within the frameworks of the orthodox model[1,2] the shape of the bottom part of $R(T)$ dependence in a 1D superconductor is associated with nucleation of a finite resistance ascribed to thermal fluctuations. Qualitatively, this behavior can be understood considering that for a sufficiently long 1D channel there is always a finite probability of a thermal fluctuation driving instantly a fraction of a wire into a normal state. As there is only one channel for supercurrent to flow, these events cause bursts of a normal current. Being integrated in time, these instant voltage jumps are experimentally observed as a finite effective resistance. The smallest volume $\Omega_\xi$ of a 1D wire with cross section $\sigma$ to be driven normal is $\Omega_\xi = \xi\sigma$. The energy required for this process is the corresponding superconducting condensation energy $\Delta F \sim B_c \Omega_\xi$, where $B_c$ is the critical magnetic field. It has been shown[1,2] that the effective resistance $R(T)$ is proportional to the probability of these events $\sim\exp(-\Delta F/k_B T)$, where $k_B$ is the Boltzmann constant. Predictions of the model[1,2] have been experimentally confirmed by measuring the shape of $R(T)$ dependencies of extremely homogeneous superconducting 1D crystals (whiskers) with effective diameter $\sqrt{\sigma} \sim 0.5$ $\mu$m.[3,4]

Formally the above process can be described as a thermally activated jump of the system from one local potential minimum into the neighboring one separated by $\pm 2\pi$ in the $\varphi$-space (so-called, phase slip). One can make a formal analogy with a "classical" jump of a particle (e.g., a gas molecule) over an energy barrier $\Delta F$ provoked by a thermal energy $k_B T$. However, taking into account that the system is quantum, an alternative mechanism has been proposed that associates the system transfer from one local potential minimum into another with tunneling through the barrier.[5] The phase difference between two remote points of the system might change in time due to quantum phase fluctuations. This purely quantum phenomenon of phase tunneling, also called as quantum phase slip (QPS), should provide an additional channel for energy dissipation in a current-carrying system. Estimations require superconducting wires with effective diameter $\sqrt{\sigma} < 10$ nm, where this quantum mechanism should dominate the "conventional" thermally activated behavior. There have been few attempts to observe the phenomenon.[6−9] In reports[6,7] Pb, In, and InPb nanowires of effective diameter $\sqrt{\sigma} > 20$ nm were studied. In recent works,[8,9] carbon nanotubes of width ∼10 nm were used as substrates supporting MoGe superconducting film. We strongly believe that the authors[6−9] did observe the claimed quantum phase tunneling behavior. Our motivation was to perform an additional experimental check eliminating uncertainty related to the uniqueness of each individual nanostructure of a certain wire diameter.[6−9] In other words, we want to trace the crossover from the thermally activated mechanism to the quantum one in the same sample, thereby evidencing for a solely size dependent origin of the phenomenon.

* Corresponding author: Phone +358-(0)14-260 2609. Fax +358-(0)-14-250 2351. E-mail Konstantin.Arutyunov@phys.jyu.fi.

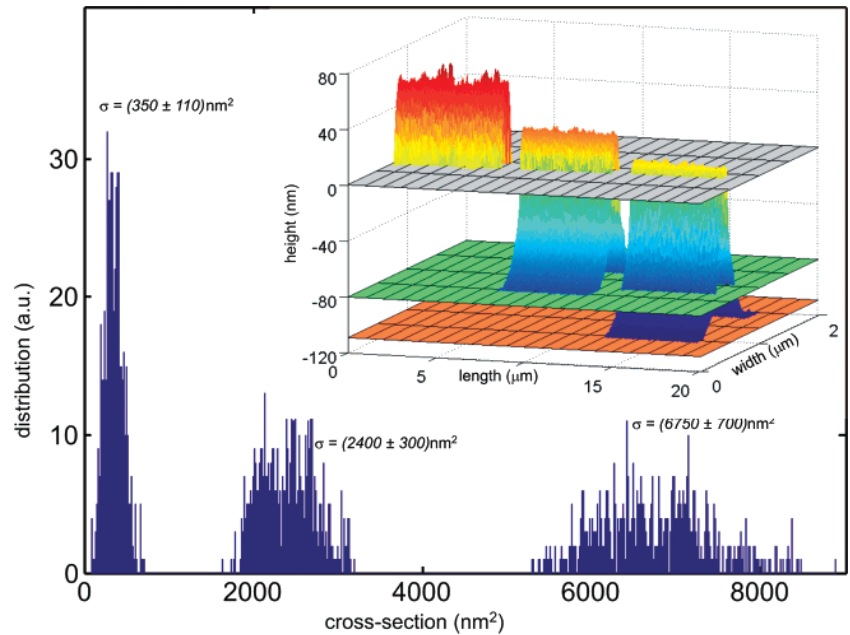

**Figure 1.** Histograms showing the distribution of the wire cross section before and after sputtering. To collect statistics, about 500 SPM scans were taken across the wire with the step along its axis ∼12 nm, which is comparable to the radius of curvature of the SPM tip. Narrowing of the histograms is due to the "polishing" effect of ion sputtering. The inset shows the evolution of the sample shape while sputtering measured by SPM. Bright color above the gray plane corresponds to Al, blue below the plane is Si. Planes (grey, green, orange) indicate Si substrate base levels after successive sessions of sputtering. As Si is sputtered faster then Al, finally the wire is situated at the top of the Si pedestal. Gray plane (height = 0) separates Si from Al.

We have applied the method of $Ar^+$ ion sputtering for progressive reduction of a nanowire effective cross section enabling measurements of the same sample between the sputtering sessions.[10] The original aluminum nanowires with typical dimensions 60 nm × 100 nm × 10 μm were fabricated on Si substrates using conventional e-beam lithography and e-gun evaporation of 99.995% pure aluminum followed by a lift-off process. An extensive scanning probe microscope (SPM) and scanning electron microscope (SEM) analysis was done to select structures without noticeable geometrical imperfections. Only the samples with no obvious artifacts on $R(T)$ dependencies were further processed. The inset in Figure 1 shows the evolution of the wire geometry with sputtering. A remarkable feature of the method is that the low energy (∼1 keV) $Ar^+$ ion bombardment provides "smooth" polishing effect: the initial roughness of a structure is reduced (Figure 1). At such low energies, $Ar^+$ ions penetrate inside the metal film at a depth of ∼1.5 nm,[11] that is comparable with the thickness of a natural oxide on aluminum surface. Thus, if the original wire has no structural defects they cannot be introduced later on by ion sputtering. The effective wire cross section $\sigma$ (Figure 2, table) has been obtained as a best-fit parameter for model calculations and appeared in a reasonable agreement with SPM and SEM measurements.[12] All experiments were performed using carefully RF filtered lines. Four-probe measurements were performed in both DC and lock-in AC modes. Temperature of the superfluid $^4$He bath was monitored by a PID controller with an accuracy ± 0.1 mK.

Figure 2 shows the $R(T)$ dependencies measured on the same Al nanowire before and after sputtering sessions. One can easily see that for larger effective cross section diameters (from $\sqrt{\sigma} \geq 17$ nm), the $R(T)$ dependencies follow the same qualitative behavior: relatively narrow transition with quasi-linear slope in logarithmic scale. These experimental data can be fitted with a reasonable accuracy by theoretical calculations within the model of thermally activated phase slips.[1,2] Note the systematic shift of the mean critical temperature $T_c$ to higher values with reduction of the wire cross section. This is a well-known experimental fact: critical temperature of superconducting films (and wires) often differs from the corresponding bulk value. For aluminum the $T_c$ always increases with the decrease of the minimum characteristic dimension. Though we believe that the $T_c$ shift phenomenon is not directly associated with the quantum phase tunneling effect, the size dependence of a superconducting critical temperature is currently under intensive discussion.[13,14]

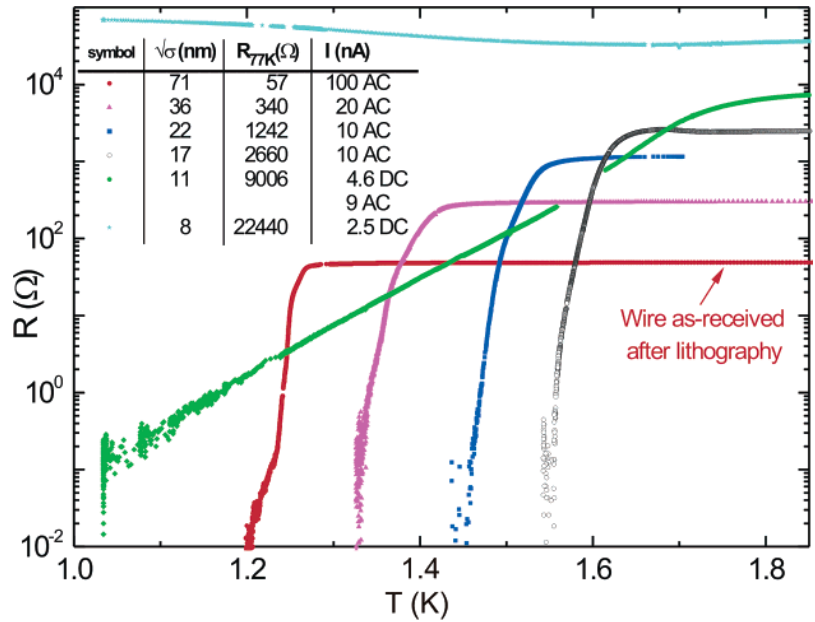

**Figure 2.** Resistance vs temperature for the *same* wire of length $L = 10\ \mu$m after several sputtering sessions. The sample and the measurement parameters are listed in the table. For low-Ohmic samples, lock-in AC measurements with the front-end preamplifier with input impedance 100 k$\Omega$ were used; for resistance above $\sim$500 $\Omega$ we used DC nanovolt preamplifier with input impedance $\sim$1G$\Omega$. The absence of data for the 11 nm sample at $T \sim 1.6$ K is due to switching from DC to AC setup. Note the qualitative difference of $R(T)$ dependencies for the two thinnest wires from the thicker ones.

When the wire effective diameter reaches $\sim$11 nm and the corresponding normal state resistance $R_N \sim 9$ k$\Omega$, the shape of the $R(T)$ dependence undergoes dramatic changes: it suddenly becomes much "wider" (Figure 2). Contrary to thicker wires, the shape of the transition cannot be fitted by the model of thermally activated phase slips[1,2] at any reasonable set of parameters (Figure 3). Even an assumption of the existence of an unrealistically narrow and long constriction within the wire (however, not confirmed by SEM or SPM) does not help. Broad $R(T)$ transition for thinner wires are assumed to be associated with the quantum phase tunneling effect or, in other words, quantum phase slippage. There are several models currently available in the literature for theoretical description of this phenomenon.[7,15−20] Calculations using phenomenological QPS approach[7] give temperature dependence $R(T)$ much weaker compared to the thermal activation model (red line in Figure 3). However, quantitatively the discrepancy between our experimental data and the model[7] predictions is about tens of orders in magnitude.

Here we present a comparison of our data with the renormalization theory.[18] Full version of the model[18] for a wire of an arbitrary length $L$ is rather sophisticated, and a comprehensive discussion would require complexity not acceptable for this type of short reports. However, if the wire is short enough that only one phase slip event occurs at a time, one can derive a simple phenomenological model utilizing the renormalization theory[18] for the rate of QPS activation $\Gamma_Q$. The requirement[18] for applicability of the "short wire" limit is that the length of the wire $L$ is to be much smaller than the characteristic scale: $L = hc_0/(k_B T) \sim 100\ \mu$m for a $\sqrt{\sigma} = 10$ nm aluminum wire with the mean free path $l = 4$ nm at $T = 1.0$ K, where $c_0 \sim 2 \times 10^6$ m/s is the velocity of the Mooij−Schön mode[21] for our "dirty" limit material.

Within the Ginzburg−Landau time $\tau_0 \sim h/\Delta$, a QPS event breaks the coherence in a section of the wire of length $\xi$, producing a voltage jump $\delta V = I(R_N/L)\xi$, where $\Delta$ is the gap parameter, $h$ is the Planck constant, $I$ is the measuring current, and $L$ is the length of the wire. The effective voltage $V_{\text{eff}}$ is measured in experiment over a much longer time $t \gg \tau_0$: $V_{\text{eff}} = \delta V \tau_0 \Gamma_Q$. The effective resistance $R_{\text{eff}}(T) = V_{\text{eff}}/I = R_N(\xi/L)(\tau_0 \cdot \Gamma_Q) = AR_Q(L/\xi) \exp(-S_{\text{QPS}})$, where $S_{\text{QPS}} = A(R_Q/\xi)/(R_N/L)$, $R_Q = h/(4e^2) = 6.47$ k$\Omega$ is the "superconducting" quantum resistance, and $A$ is a numerical coefficient $\sim 1$.[18] Contrary to thermally activated PS,[1,2] the QPS contribution has a rather weak temperature dependence determined by the temperature variation of the coherence length $\xi(T)$, and should produce a finite resistance even at $T \rightarrow 0$ for sufficiently narrow wires. The comparison of the calculations with our data is presented in Figure 3. Taking into consideration possible limitations of the "short wire" approach, the correspondence between the simplified version of the model[18] and the experimental data is rather good. Probably, the discrepancy observed at low temperatures (Figure 3) can be removed provided that the full version of model[18] is applied.

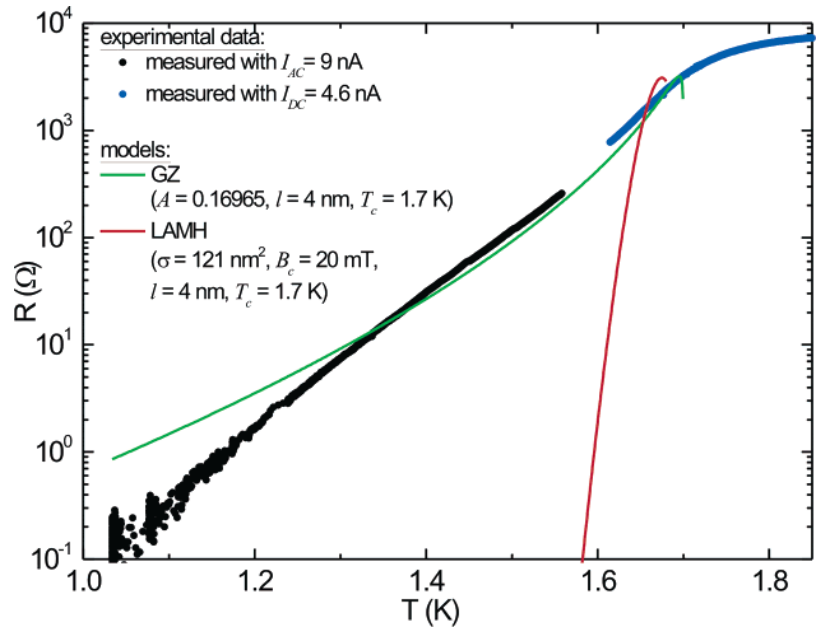

**Figure 3.** $R(T)$ dependence for the $\sqrt{\sigma} \sim 11$ nm sample. Green line shows the result of fitting to renormalization theory[18] with $A$, $l$, and $T_c$ being fitting parameters. The same set of parameters together with the critical magnetic field $B_c$ measured experimentally is used to show corresponding effect of thermally activated phase slips on the wire's $R(T)$ transition (red line).[1,2] The parameter $\sigma$ is obtained from the normal state resistance value and the known sample geometry.[12] The estimation for $\sigma$ is in a reasonable agreement with SPM analysis as well as with evolution of $\sigma$ over all sputtering sessions (see Figure 2).

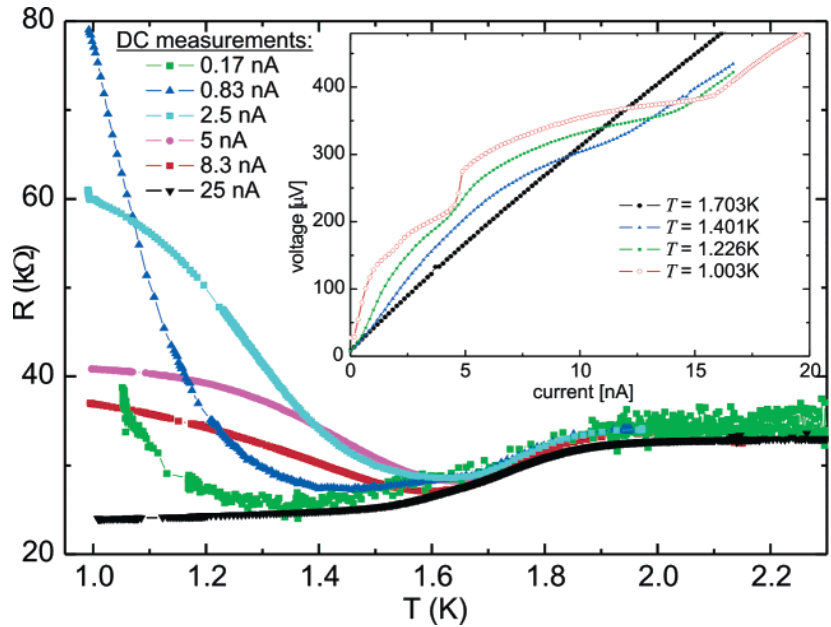

**Figure 4.** $R(T)$ dependencies for the $\sqrt{\sigma} \sim 8$ nm sample. Note the unusual nonmonotonic dependence of $R(T, I = \text{const})$ on the measuring current $I$. The inset shows $V(I)$ characteristics taken at several temperatures. At lower temperatures one can clearly observe the development of a nonlinearity.

A remarkable outcome of the model[18] is that for a given sample of length $L$ the total resistance in the normal state $R_N$ is of critical importance. Assuming the parameter $A \sim 1$, the QPS mechanism starts to be exponentially pronounced as soon as the strong inequality $(R_N/L) \ll (R_Q/\xi)$ is no longer valid. We indeed did observe the qualitative change in the shape of $R(T)$ dependencies when the normal state resistance of the wire of the same length increased from $\sim 2.6$ k$\Omega$ to 9 k$\Omega$ (Figure 2). With further reduction of the wire cross section (Figures 2 and 4), the metallic behavior turns into semiconducting: resistance increases with decreasing temperature from 77 K down to 2 K. Below the expected critical temperature $\sim 1.7$ K the shape of the $R(T)$ curve strongly depends on the measuring current: from superconducting to insulating behavior (Figure 4). The $V(I)$ characteristics (Figure 4, inset) show a nonmonotonic dependency at lower temperatures and Ohmic behavior close to the critical temperature. Contrary to this thinnest wire, the thicker samples display a pronounced single step in $V(I)$ characteristics at temperatures below the critical with a well-defined critical current $I_c \sim T^{3/2}$. The latter confirms 1D nature of the studied systems.[22,23]

When the normalized normal state resistance $R_N/L$ of a wire increases, the rate of quantum fluctuations $\Gamma_Q$ might become so high that the true zero resistance cannot be obtained even at $T \rightarrow 0$. We believe that the thinnest wire $\sqrt{\sigma} \sim 8$ nm and $R_N \sim 33.4$ k$\Omega$ is already in this limit. The nonmonotonic behavior of $R(T, I = \text{const})$ on measuring current $I$ is rather unusual (Figure 4). The corresponding $V(I)$ characteristics taken below the expected critical temperature (Figure 4, inset) show deviations from Ohm's law, which might be associated with critical current. Neither hysteresis, nor S-shape $I-V$ characteristics[24] have been observed in our experiments. The wire is so thin that, though its homogeneity has been studied with SPM, one might consider that tiny tunnel barriers can be formed somewhere along the wire. Semiconducting behavior of the wire resistance observed below 77 K might be ascribed to formation of such superconductor-insulator-superconductor (S−I−S) junctions. However, in this case the subgap current is surprisingly high and the voltage peculiarity corresponds to $\Delta/e$ rather than to $\Delta/2e$ typical for N−I−S systems. The absence of a Coulomb blockade signature on $I-V$ characteristics at temperatures above the critical is also rather confusing. Thus, trivial formation of S−I−S junctions in the thinnest ∼8 nm wire cannot account for the peculiarities in the $I-V$ characteristics in Figure 4. On the other hand, one might think about a "tricky" combination of a S−I−S junction of two QPS dominated 1D "superconducting" contacts separated by a thin tunnel barrier.

In summary, we have observed an evolution of the shape of $R(T)$ dependencies in ultrathin and homogeneous aluminum nanowires as a function of their effective diameter. With a high level of confidence, the artifacts related to inhomogeneity of the wires can be ruled out, except, probably, for the thinnest sub-10 nm wires. With these reservations, we believe that in our experiments we did observe a size effect. It has been found that for effective diameters $\sqrt{\sigma} \geq 15$ nm the $R(T)$ dependencies can be described by the model of thermally activated phase slips.[1,2] For thinner wires the width of $R(T)$ transition is dramatically wider, and the resistance may not tend to zero at $T \rightarrow 0$. We associate this phenomenon with manifestation of quantum phase fluctuations (quantum phase slips). The effect should have a universal validity for all superconducting materials. Apart from Al, other materials such as Sn, In, Pb are on the list of our current research. The study of quantum fluctuations phenomenon suppressing superconductivity in ultranarrow wires is of great importance for future development of nanoelectronics. It puts a fundamental limitation on utilization of superconducting nanocomponents designed to transport a dissipationless electric current.

**Acknowledgment.** The authors would like to acknowledge D. Golubev and A. Zaikin for their helpful discussions. The work was supported by the Academy of Finland under the Finnish Center of Excellence Program 2000-2005 No. 44875, Nuclear and Condensed Matter Program at JYFL, and the EU Commission FP6 NMP-3 project 505457-1 ULTRA-1D "Experimental and theoretical investigation of electron transport in ultranarrow 1-dimensional nanostructures".